\documentclass{elsart}
\usepackage{amssymb}
\usepackage{amsmath}
\usepackage{graphics}

\begin{document}
\begin{frontmatter}
\title{\large \bf On  contribution of three-body forces  to  $Nd$ interaction
 at intermediate energies
}

\author[dubna]{Yu.N.~Uzikov}
\address[dubna]{JINR, LNP, Dubna, 141980, Moscow Region, Russia}

\begin{abstract}
{ Available data on  large-angle nucleon-deuteron 
 elastic scattering $Nd\to dN$  below the pion threshold 
 give a signal for three-body forces.
 There is a  problem of separation of possible subtle
 aspects of these forces  from off-shell effects in  two-nucleon potentials.
 By considering the main mechanisms  of the process $Nd\to dN$,
  we show  qualitatively
 that in the quasi-binary reaction
 $N+d\to (NN)+N$  with the final spin singlet NN-pair in the  S-state
 the relative contribution of the 3N forces  differs substantially
 from the  elastic channel. It gives a new testing ground
 for the problem in question.}

\end{abstract}
\vspace{5mm}
\vspace{8mm}
\noindent
\begin{keyword}
 proton deuteron  scattering,
 three nucleon forces
\begin{PACS}
24.70.+s; 24.50.+g; 21.45.+v\\[1ex]
\end{PACS}
\end{keyword}
\end{frontmatter}

  An existence of the three-nucleon forces (3NF) is not doubted  both in the 
 standard  meson-exchange picture \cite{coon79} and in the
 ciral perturbation theory  \cite{friar99}. 
 Their strength and detail structure are still under discussion
 \cite{kievsky99}.
 An often used  model of a 3NF  is the $2\pi$ exchange
 in the form called the Tucson-Melbourne (TM-3NF) parametrization 
 \cite{coon79} (Fig.1).
  At present nucleon-nucleon (NN)  forces underestimate
  binding energies of the light
  nuclei \cite{carlson98}.  The 3NF allow one  to fill in part the gaps. 
 Another  signal for the 3NF gives  elastic $Nd$
  scattering  below  200 MeV.
  Recent experimental and theoretical investigations 
 \cite{witala94,witala98,witala99,cadman00,bieber00}
  show  that  reach set of spin observables in this process
  gives a real chance to
  study various aspects of 3NF effects. So, the measured cross section,
  which is underestimated at the scattering angles
 $\theta_{cm}=60^\circ-180^\circ$
 by the  data-equivalent modern NN forces, 
 is excellently reproduced  by three-body calculations   with 
 the TM-3NF. The deuteron analyzing power $A_y^d$ is also well reproduced
 \cite{sakai}.  On the other hand,
 the nucleon analyzing power $A_y^p$ is still under discussion
 \cite{hueber98},  \cite{canton}. Furthermore,
 the same approach fails to explain
 tensor analyzing powers from the precise $dp$ scattering data
 \cite{sakai}. From this observation the authors of
 Ref. \cite{sakai} conclude
 that there is  deficiencies in the spin structure of the TM-3NF.
  
  However, one should note that 
  off-shell properties  of NN forces contribute simultaneously with 
  the 3NF. The problem is that these ambiguous
   aspects of NN forces can be studied  in the 3N or many nucleon systems only
  (see recent  discussion in Ref. \cite{canton}). A noticeable NN input
  dependence was observed in Refs.\cite{sakai,witalpl99}.
  In  this connection
  the authors of  Ref. \cite{witalpl99} proposed to measure the zero point
  of the longitudinal asymmetry in the ${\vec n}{\vec d}$ total cross section
  in order to get  more clear signal  for 3NF effects.  This letter is
  another step in that direction.
  Namely, to separate more definitely  the NN and 3NF effects in the
  elastic Nd  scattering  one should  investigate   supplementary   
  processes  at almost the same kinematics
  but with a different relative role of the  NN and 3NF.
  As  shown here, a substantial different NN-to-3N ratio occurs
  in  the $pd$-interaction
  with formation of the spin-singlet NN-pair in the final $^1S_0$
  state.  We propose here to study  the  reactions
\begin{equation}
\label{r1}
  p+d\to (pp)+n 
\end{equation}
   and  $n+d\to (nn)+p$ 
  at  large cm scattering  angle of the secondary nucleon and 
  low relative energy of two protons $(pp)$ or neutrons $(nn)$
  $E_{NN}<3$ MeV, when the $^1S_0$ state dominates.
  The reaction (\ref{r1}) was not yet investigated experimentally.
  Complete theoretical analysis of this reaction is possible 
  at present at energies
  below the pion threshold (214  MeV)  in the framework of
  rigorous 3N-scattering  approaches \cite{gloekle96,viviani98}.
  As shown  by Faddeev calculation,
   at initial energies below about 200 MeV
  rescattering of higher order is very important.  However, around 300 MeV
  the first two terms in the multiple scattering expansion  are sufficient
  to describe the total $nd$ cross section \cite{witala99}. Therefore, 
  the first Born approximation
  can be used  as a qualitative estimation near  the pion threshold.
  Within this approximation one cannot find an exact contribution of the 3NF. 
  Nevertheless, we can  show  qualitatively that the
  relative role of the NN and  3NF forces   differs 
  in this reaction  considerably from that in the elastic Nd scattering.  
  First, using isospin invariance, we  show  that the Born 3NF amplitude 
  (Fig.1a) of the reaction with the singlet $pn(^1S_0)$ pair
  is suppressed by the factor of $\frac{1}{3}$ in respect to the 3NF amplitude
  with the  deuteron. On the other hand, the Born term of the one nucleon
  exchange (ONE)  mechanism (Fig.1c), related to  the NN-forces,
  is not affected by isospin factors.  Second, the ONE 
  contribution is  modified  considerably
  due to suppression of the higher orbital momenta $l\not =0$ in the
  final NN-system  at low $E_{NN}$ whereas it is not so for the 3NF.
  In additional, an important modification of 
  the single scattering (SS) mechanism (Fig.1d)
  occurs  in the  reaction (\ref{r1}) in comparison with the $pd\to dp$.

 The dynamics of the reaction (\ref{r1}) is discussed here
 by analogy with known mechanisms of the backward elastic pd-scattering
 \cite{lak,bd} (see Fig.1).
  As we show below, the ONE+SS sum dominates  in the
  $pd\to dp$ process  at kinetic  energy of the proton beam
  $T_p\sim 0.2-0.3$ GeV.
  At higher energies $T_p=0.4-1.0$ GeV  the double $pN$-scattering 
  (Fig.1e) with the $\Delta$-excitation ($\Delta$) gives the main contribution
 \cite{lak,uz98}. 
 For the NN$\rightleftharpoons \Delta$N  amplitude we use here
 the $\pi+\rho$  exchange model \cite{imuz88} depicted in Fig.1b.
 This model describes the measured cross section  of
 the pp$\to pn\pi^+$ reaction \cite{hg} in the $\Delta$-region.
 Since the $\Delta$ mechanism is an important  ingredient of
 the 3NF amplitude \cite{nemoto}, we identify here the $\Delta$ 
 contribution  with the 3NF one. The $\Delta$-isobar is considered
  as a stable baryon, that is appropriate  below the pion production
 threshold. 

 Let us discuss  the isotopic spin factors
 for the Born 3NF term of the   $p\,d\to (np)_{s,t}\,p$
 amplitude depicted in Fig.1a. The $2\pi$ exchange mechanism
 contains  two terms corresponding to different values
 of the total isospin  of the intermediate meson-nucleon  system,
 $T=\frac{1}{2}$ and $\frac{3}{2}$:
\begin{eqnarray}
\label{isot3}
A(p\,d\to (pn)_s\,p) = { A}^{s}_{T=1/2}+ {A}^{s}_{T=3/2}, \\ \nonumber
A(p\,d\to (pn)_t\,p) = { A}^{t}_{T=1/2}+ { A}^t_{T=3/2},
\end{eqnarray}
 where the spin singlet ($s$) and triplet ($t$) states of
 the final pn-pair correspond to  the isospin $T_{pn}=1$ and $0$,
 respectively. 
 The  isospin  structure of the  amplitudes ${A}^{s,t}_{T}$ 
 is given by
$$A_{T}^{s,t}(p\,d\to (pn)\,p)=(2T+1)\sqrt{2(2T_{pn}+1)}
(T_{pn}0\frac{1}{2} \frac{1}{2}|\frac{1}{2} \frac{1}{2})\times$$
\begin{equation}
\label{isot3a}
\times\left \{ \begin{array} {ccc} 1&\frac{1}{2}&\frac{1}{2}\\
 1&\frac{1}{2}& T
\end{array}\right \}
\left \{ \begin{array} {ccc} \frac{1}{2}&\frac{1}{2}&T_{pn}\\
 \frac{1}{2}&\frac{1}{2}&1
\end{array}\right \} B_T^{s,t}.
\end{equation}
 The Clebsh-Gordan coefficients and 6j-symbols are used here in
standard notations.
The dynamical factor $B_T$ in Eq. (\ref{isot3a}) does not depend
on  z-projections of the isotopic spins.
Assuming the spatial parts of the singlet and triplet
wave functions of the 
$pn$ pair to be the same, i.e. $B_T^s=B_T^t$,
  one can find
from Eq.(\ref{isot3a}) the following ratios
\begin{eqnarray}
\label{isot4}
r=\frac{{ A}^{s}_{T=1/2}}{{ A}^{t}_{T=1/2}}=
\frac {{ A}^{s}_{T=3/2}}{{ A}^{t}_{T=3/2}}= -\frac{1}{3}.
\end{eqnarray}
 After substituting  Eq. (\ref{isot4}) into Eq. (\ref{isot3}),
 one finds 
 \begin{equation}
\label{isot6}
 R_I^{iv}=\frac{A(p\,d\to (pn)_s\, p)}{A(p\,d\to (pn)_t\, p)} =- \frac{1}{3}.
  \end{equation}

 We stress that owing to Eq.(\ref{isot4})
 the ratio (\ref{isot6}) does not depend on the unknown
 relative phase nor on the ratio of the amplitudes
 $A_{1/2}$ and  $A_{3/2}$ of the virtual process $\pi N\to \pi N$ in
 Eq.(\ref{isot3}). The result given by Eq.(\ref{isot6}) is valid not only
 for the $\Delta$-mechanism (Fig.1e), as  was found in Ref. \cite{imuz87}.  
 In fact,  all intermediate states
 of the  meson-nucleon  system both  for the isotopic spin
 T=3/2 and  T=1/2 are taken into account in Eq.(\ref{isot6})
 including the $\Delta$ and  $N^*$ poles  and  the $\pi N$ continuum.
 Obviously,  the relation (\ref{isot6}) is valid also
 for the sum of the  diagrams in Fig.1a
 with   different combinations of the   isovector mesons
 ($\pi \pi, \pi\rho, \rho \rho , \dots $), as well as for
  the  reaction $pd\to dN^*$.

 For the isoscalar  meson exchange
 ($\omega, \ \eta,\ \eta ',\, \dots $)   we find the ratio
 $R_I^{is}=1$. The same ratio is valid  for the ONE mechanism,
 $R_I^{ONE}=1$. It is impossible to write  a definite 
 isotopic factor for the SS-mechanism because
 in this case the $s/t$ ratios are different for
 the isoscalar ($r=1$) and isovector ($r=\frac{1}{3}$) NN-amplitudes,
 which are mixed with an unknown relative phase in the upper vertex of
 the diagram  in Fig.1d.

  A detail formalism for the amplitude of the reaction $pd\to (NN)N$
 in the framework of the ONE+SS+$\Delta$ model can be derived from the 
  $pd\to dp$ formalism of Refs. \cite{lak,uz98}. For this 
 aim one should  make the following substitution into the matrix elements
 \begin{equation}
 \label{sub}
 |\varphi_d>\to \sqrt{m}|\Psi_{ k}^{(-)}>,
 \end{equation}  
 where $|\varphi_d>$ is the deuteron final state in the  $pd\to dp$,
 $m$ is the nucleon mass  and
 $ |\Psi_{ k}^{(-)}>$  is the scattering state of the final
 NN-system  at the relative momentum ${ k}$
 in the $N+d\to(NN)+N$ reaction.
 Since  the S-wave gives the main contribution to the singlet
  $(NN)_s$-state at  
 $E_{NN}=\frac{k^2}{m}< 3$ MeV, one should omit the D-component of
 the final deuteron state
 $|\varphi_d>$ in the $pd\to dp$  formalism  \cite{lak,uz98}
 when making the substitution (\ref{sub}).
  Thus,  one has to insert into the  upper vertex of the ONE diagram (Fig.1c)
  the half-off-shell amplitude of $NN$-scattering in the $^1S_0$ state,
  $t_s(q,k)$.
 This amplitude,  as a function
 of the off-shell momentum $q$, is very close in  shape 
 to the deuteron S-wave function
 in momentum space, $u(q)$, and  
 has  a node at the point $q\sim 0.4$ GeV/c.
 The node is caused by the short-range repulsion
 in the $NN$ potential. A similar node available in the
 wave function  $u(q)$  can be  connected  to the
 null  of  the measured deuteron charge formfactor $G_{C}(Q)$ 
 at the  transferred momentum $Q\sim 4.5$ fm$^{-1}$ \cite{t20}.  The 
 node of $u(q)$  was not yet observed directly in any  reactions with
 the deuteron  due to a large  contribution of the deuteron D-state. 
 An important feature of the reaction
 (\ref{r1}) is a possibility to display  the node
 of the amplitude $t_s(q,k)$ straightforwardly in the cross section
   at $T_p=600-700$ MeV and $\theta_{cm}=180^\circ$
 \cite{imuz90,smuz98}. At initial energies 100-300 MeV this node makes
 the ONE contribution  vanishing at $\theta_{cm}=100^\circ-130^\circ$.

  The results of our calculations  performed within the  ONE+SS+$\Delta$ 
 model with the Paris NN-potential  are shown in Fig.2 for
 the $pd\to dp$ process and  in Fig.3 for the reaction (\ref{r1}).
 The model describes rather well the $pd \to dp$ cross section 
 at $T_p=150-250$ MeV and $\theta_{cm}>120^\circ$.
 The $\Delta(\equiv $3NF) contribution strongly depends on the cutoff
 parameters
 $\Lambda_{\pi,\rho}$ in the NN$\rightleftharpoons \Delta$N  amplitude.
  We use here 
 the values 
 $\Lambda_\pi=0.6$ GeV and  $\Lambda_\rho=0.7$ GeV obtained from the 
 fit the data on $pp\to pn\pi^+$ and $pd\to dp$ \cite{imuz88,uz98}.
 The sum ONE+SS underestimates the cross section at
 $\theta_{cm}=110^\circ-130^\circ$, but this discrepancy is eliminated
 by adding the $\Delta$-contribution (Fig.2a), as was observed in 
 Refs.\cite{witala98,nemoto}. The calculated analyzing powers $A^p_y$,
 $A^d_y$, and  $A_{yy}$ are  in only qualitative agreement with the data 
 (Fig.2b-d).
 
 In contrast to the $pd\to dp$ process,
 the influence  of the $\Delta$  mechanism 
 in the  reaction (\ref{r1}) is rather
 weak in  the cross section, but more pronounced in the analyzing powers
 (Fig.3). 
 At the minimum of the  cross section, $\theta^n_{cm}=120^\circ-140^\circ$,
 the role of 3NF increases due to i) vanishing of the ONE amplitude
 and ii) rather fast decreasing of the SS contribution (Fig.3a).
  Outside of this region the $\Delta$ contribution to the reaction
 (\ref{r1}) is  smaller than  in the elastic $pd$ scattering owing to 
 the isospin relations. 
 Note,  within the  ONE+SS approximation the behaviour of the vector
 analyzing powers   $A^p_y$ and 
 $A^d_{y}$ is considerably  different in the reaction (\ref{r1})
 as compared  to the $pd\to dp$ process (Figs. 3b,d).
 The reason   is a modified  structure of the SS-mechanism. 
 Indeed, only the  pn-scattering at small angle contributes to the
 upper vertex of the SS  mechanism in the reaction (\ref{r1}) 
 \cite{imuz90}. On the contrary,  both  the charge exchange process
 $pn\to np$ and the $pp$ elastic scattering at small angles
 contribute to the  $pd\to dp$ process \cite{lak}.
 The $\Delta$ mechanism taken into account in addition to the 
 ONE+SS sum, changes the  analyzing powers 
 noticeably (Fig.3c,d)

 In conclusion, the ONE+SS+$\Delta$ model
allows one to  understand 
 qualitatively
 the main features of the $pd\to dp$ observables at $T_p \sim 200$ MeV.
 Within this model we found  that the relative contribution of the 3NF
 in the  reaction $N+d\to (NN)(^1S_0)+N$ 
 differs considerably from the  elastic $pd$ scattering. 
 A sizable  modification of the  analyzing powers 
 is expected in the  reaction
 (\ref{r1}) in comparison with  the $pd\to dp$, in particular,
 due to the 3NF effects.  Future experimental study 
 of the reaction (\ref{r1}) near the pion threshold and 
  rigorous three-body calculations,  complementing 
 recent analysis of
 the process $pd\to dp$, can gain more insight into the 3NF.


\input epsf
\begin{figure}
\vspace*{-1cm}
\mbox{\epsfxsize=6.3in \epsfbox{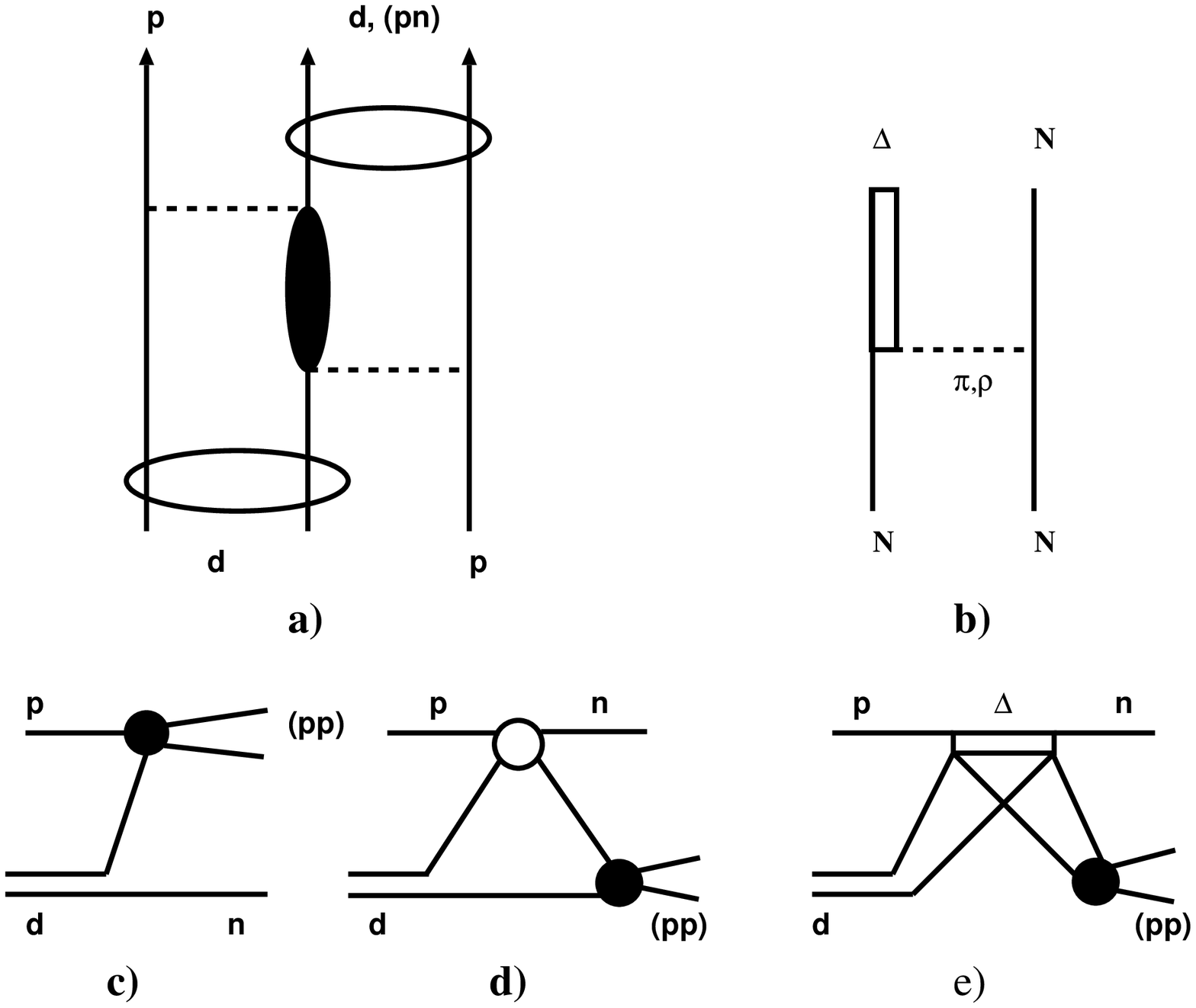}}
\caption{ Mechanisms of the reaction $p+d\to (NN)+N$:
   $a)$ --  the  Born 3NF amplitude of the 
  $pd\to (pn)p$ and $pd\to pd$ processes,
   $ b)$ ---   $\pi+\rho$ exchange for the 
    $NN\to N \Delta$ amplitude; 
       $ (c) $  --  one nucleon exchange,
       $ (d)$ --   single scattering,
       $ (e)$ --   $\Delta$-isobar excitation
}
\label{f1}
\end{figure}

\eject
\begin{figure}
\begin{center}
\vspace*{-1cm}
\mbox{\epsfxsize=7.0in \epsfbox{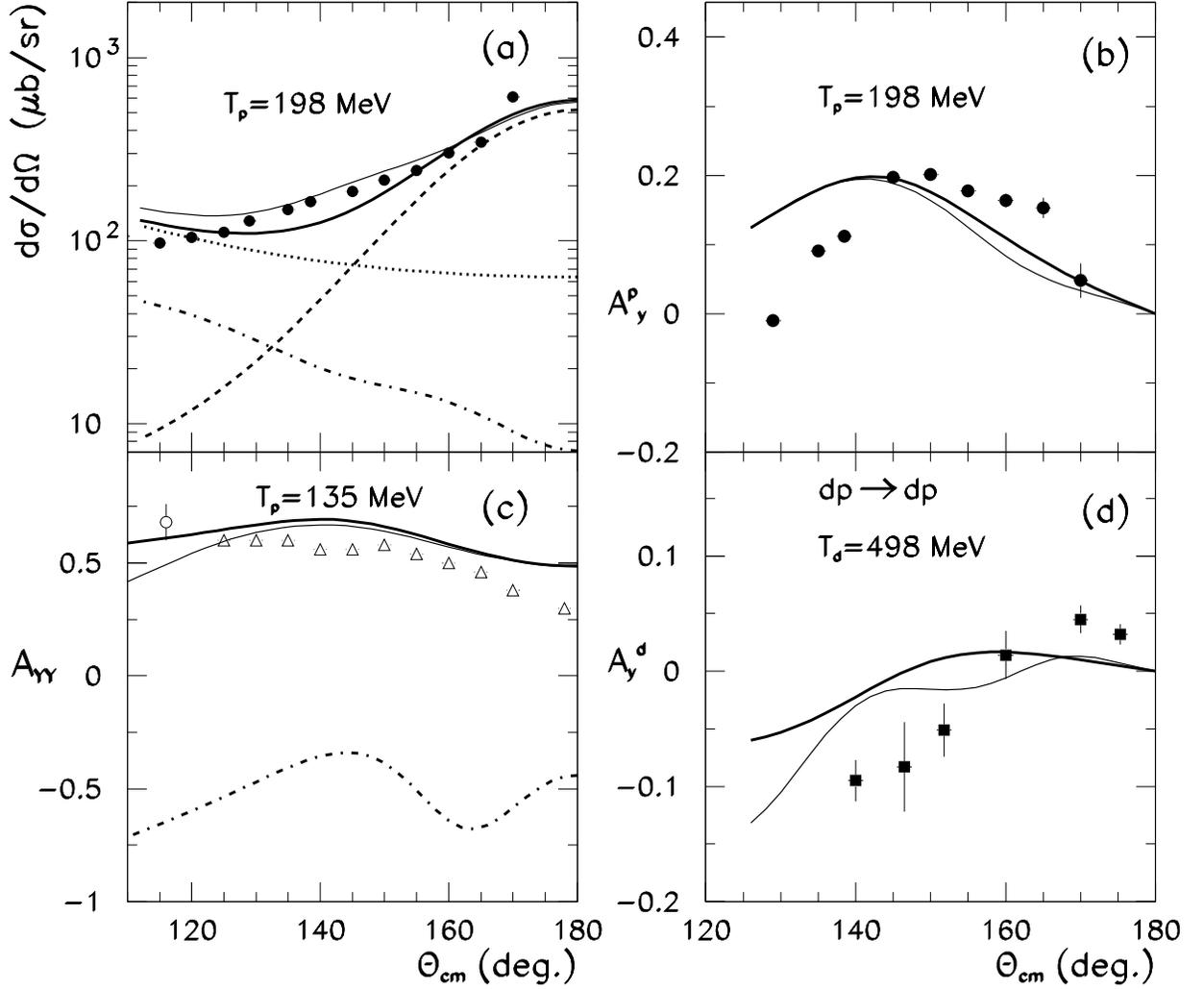}}
\caption{The cm cross section ({\it a}) and analyzing powers
 $A_y^p$ ({\it b}),
 $A_y^d$({\it d}),  $A_{yy}^d$({\it c}) in the 
 pd-elastic scattering in backward hemisphere at different initial
 energies of the proton $T_p$ and the deuteron $T_d$ ($T_p=\frac{1}{2}T_d$).
 The results of 
 calculations with  the different mechanisms are compared
 with the experimental data from Refs.
 \cite{adel}($\bullet$), \cite{arv} (filled squares), and 
 \cite{sakai} (open triangles and circles): ONE (dashed line),
 SS (dotted), $\Delta$ (dashed-dotted),
 ONE+SS (full thick),  ONE+SS+$\Delta$ (full thin)
}
\label{f2}
\end{center}
\end{figure}

\eject
\begin{figure}
\begin{center}
\vspace*{-1cm}
\mbox{\epsfxsize=7.0in \epsfbox{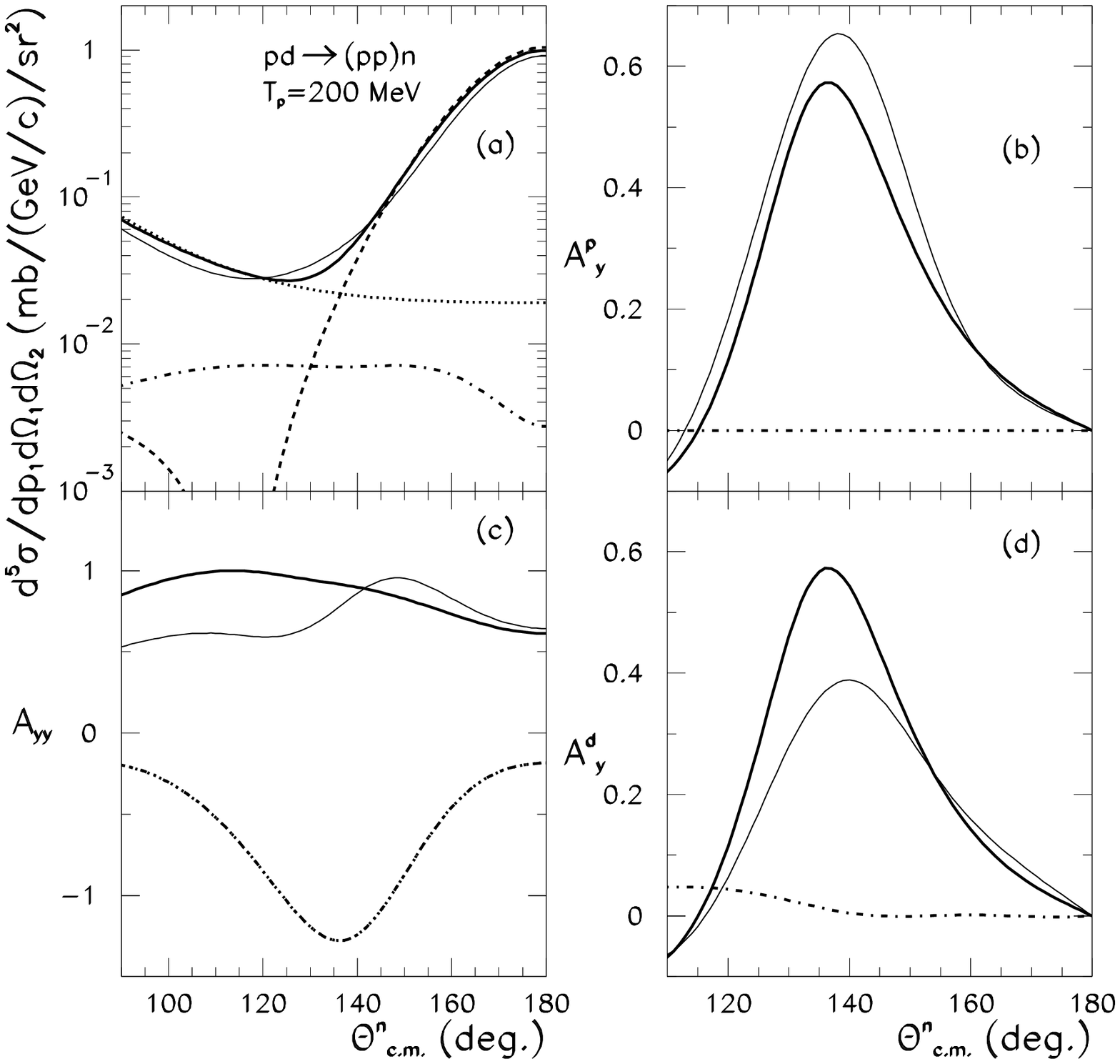}}
\caption{The same as in Fig.2, but for the reaction 
  $p+d\to (pp)_s+n$ at $T_p=200$ MeV  and relative energy of two protons
$E_{pp}=3$~MeV versus the neutron scattering angle
$\theta^n_{cm}$ 
}
\label{f3}\end{center}
\end{figure}

\begin{thebibliography}{99}
\bibitem{coon79} S.\,A. Coon, M.\,D. Scadron, P.\,C. McNamee  et al.,
 Nucl.~Phys. {\bf A317}, 242 (1979).
\bibitem{friar99} L.\,J. Friar, D. H\"uber, U.\,van Kolck,
 Phys.~Rev.  {\bf C 59}, 53 (1999).
\bibitem{kievsky99}  A. Kievsky, Phys.~Rev. {\bf C 60}, 034001 (1999).
\bibitem{carlson98} J. Carlson, R. Schiavilla, Rev.~Mod.~Phys. {\bf 70}, 
743 (1998).
\bibitem{witala94} H. Witala, D. H\"uber, W. Gl\"ockle, Phys.~Rev. {\bf C 49},
 R14 (1994).
\bibitem{witala98}  H. Witala, W. Gl\"ockle, D. H\"uber at al.,
 Phys.~Rev.~Lett. {\bf 81}, 1183 (1998).
\bibitem{witala99} H. Witala, H. Kamada, A. Nogga et al.,
 Phys.Rev. {\bf C 59}, 3035 (1999).
\bibitem{cadman00} R.\,V. Cadman, J. Brack, W.\.J. Cummings et al.,
nucl-ex/0010006.
\bibitem{bieber00} R. Bieber, W. Gl\"ockle, J. Golak et al., Phys.~Rev.~Lett.
{\bf 84}, 606 (2000)
\bibitem{sakai} H. Sakai, K. Sekiguchi, H. Witala et al., 
Phys.~Rev.~Lett. {\bf 84}, 5288 (2000).
\bibitem{hueber98} D. H\"uber, J.\,L. Friar. Phys.~Rev. {\bf C 58}, 
674 (1998).  
\bibitem{canton} L. Canton, W. Schadow, Phys.~Rev. {\bf C 62}, 044005 (2000);
 ibid  {\bf C 64}, 031001 (2001).
\bibitem{witalpl99} H. Witala, W. Gl\"ockle, J. Golak et al., Phys.~Lett.
{\bf B447}, 216 (1999).
\bibitem{gloekle96} W. Gl\"ockle, H. Witala, D. H\"uber et al., Phys.~Rep.
{\bf 274}, 107 (1996).
\bibitem{viviani98} M. Viviani, Nucl.~Phys. {\bf A 631}, 111c (1998). 
\bibitem{lak}  L.\,A. Kondratyuk, F.\,M. Lev, L.\,V. Shevchenko, 
  Yad. Fiz. {\bf 33}, 1208 (1981).
\bibitem{bd} A. Boudard, M. Dillig, Phys.~Rev. C {\bf 31}, 302 (1985).
\bibitem{uz98}  Yu.\,N. Uzikov, El.~Chast.~At.~Yadr.  {\bf 29}, 1405 (1998).
\bibitem{hg} J.Hudomaly-Gabitzsch  et al.,  Phys.~Rev.
 {\bf C18}, 2666 (1978).
\bibitem{imuz88} O. Imambekov,  Yu.\,N. Uzikov,  Yad.~Fiz.
 {\bf 47}, 1089 (1988).
\bibitem{nemoto} S.~Nemoto, K.~Chmielevsky, S.~Oryu  et al.,
 Phys.~Rev. C {\bf 58}, 2599 (1998).
\bibitem{imuz87} O. Imambekov, Yu.\,N. Uzikov, 
 Izv.~AN~SSSR, ser.~fiz. {\bf 51}, 947 (1987).
\bibitem{t20} D. Abbott, A. Ahmidouch, H. Anklin et al.,
 Phys.~Rev.~Lett. {\bf 84}, 5053 (2000).
\bibitem{imuz90} O.Imambekov,  Yu.N.Uzikov, 
 Yad.Fiz. {\bf 52}, 1361 (1990).
\bibitem{smuz98} A.\,V. Smirnov, Yu.\,N. Uzikov,  Yad.~Fiz.
 {\bf 61}, 421 (1998).
\bibitem{adel} R.\,E. Adelberger, C.\,N. Brownet, Phys.~Rev. {\bf D5},
 2139 (1972).
\bibitem{arv} J. Arvieux, S.\,D. Baker, R. Beutrey et al., Nucl.~Phys.
 {\bf A431}, 613 (1984). 
\end{thebibliography}
\end{document}